\documentclass[aip,amsmath,amssymb,reprint,twocolumn]{revtex4-1}
\usepackage[english]{babel}
\usepackage[utf8]{inputenc}
\usepackage[colorinlistoftodos, color=green!40, prependcaption]{todonotes}
\usepackage{amsthm}
\usepackage{mathtools}
\usepackage{physics}
\usepackage{xcolor}
\usepackage{graphicx}
\usepackage[left=23mm,right=13mm,top=35mm,columnsep=15pt]{geometry} 
\usepackage{adjustbox}
\usepackage{placeins}
\usepackage[T1]{fontenc}
\usepackage{lipsum}
\usepackage{csquotes}
\usepackage[pdftex, pdftitle={Article}, pdfauthor={Author}]{hyperref} % For hyperlinks in the PDF
\usepackage{graphicx}
\usepackage{gensymb}
\usepackage{changepage}
\usepackage{booktabs}

\begin{document}
\title{Persistent spin texture preserved by local symmetry in graphene/WTe$_2$ heterostructure}

\author{Przemys{\l}aw Przybysz}
\email[Correspondence should be addressed to: ]{p.p.przybysz@rug.nl or jagoda.slawinska@rug.nl}
\affiliation{Zernike Institute for Advanced Materials, University of Groningen, Nijenborgh 3, 9747 AG Groningen, The Netherlands}
\affiliation{Faculty of Physics and Applied Informatics, University of {\L\'o}d{\'z}, Pomorska 149/153, 90-236 {\L\'o}d{\'z}, Poland}

\author{Karma Tenzin}
\affiliation{Zernike Institute for Advanced Materials, University of Groningen, Nijenborgh 3, 9747 AG Groningen, The Netherlands}
%\affiliation{Department of Physical Science, Sherubtse College, Royal University of Bhutan, 42007 Kanglung, Trashigang, Bhutan}

\author{Berkay Kilic}
\affiliation{Zernike Institute for Advanced Materials, University of Groningen, Nijenborgh 3, 9747 AG Groningen, The Netherlands}

\author{Witold Koz{\l}owski}
\affiliation{Faculty of Physics and Applied Informatics, University of {\L\'o}d{\'z}, Pomorska 149/153, 90-236 {\L\'o}d{\'z}, Poland}

\author{\linebreak Pawe{\l} Kowalczyk}
\affiliation{Faculty of Physics and Applied Informatics, University of {\L\'o}d{\'z}, Pomorska 149/153, 90-236 {\L\'o}d{\'z}, Poland}

\author{Pawe{\l} Dabrowski}
\affiliation{Faculty of Physics and Applied Informatics, University of {\L\'o}d{\'z}, Pomorska 149/153, 90-236 {\L\'o}d{\'z}, Poland}

\author{Jagoda S{\l}awi{\'n}ska}
\email[Correspondence should be addressed to: ]{p.p.przybysz@rug.nl or jagoda.slawinska@rug.nl}
\affiliation{Zernike Institute for Advanced Materials, University of Groningen, Nijenborgh 3, 9747 AG Groningen, The Netherlands}

%\date{\today}

\begin{abstract}
Crystal symmetries in solids give rise to spin–momentum locking, which determines how an electron’s spin orientation depends on its momentum. This relationship, often referred to as spin texture, influences both charge-to-spin conversion and spin relaxation, making it one of the essential characteristics for spin-orbit-driven phenomena. Materials with strong spin-orbit coupling and broken inversion symmetry can host persistent spin textures (PSTs) - unidirectional spin configurations in momentum space, supporting efficient charge-to-spin conversion and extended spin lifetimes. Monolayer WTe$_2$, a topological material crystallizing in a rectangular lattice, is a notable example; its symmetry enforces a canted PST enabling quantum spin Hall effect with the nontrivial spin orientation. Here, we use first-principles calculations to explore how these properties are modified when WTe$_2$ is interfaced with graphene. We find that the PST is preserved by the local symmetry present in different regions of the heterostructure, while the system develops extended electron and hole pockets, resulting in semimetallic behavior. Although the band gap closes and eliminates the quantum spin Hall phase, spin Hall effects remain robust in both conventional and unconventional geometries. The computed spin Hall conductivities are comparable to those of other two-dimensional materials, and the survival of the PST suggests the possibility of long-range spin transport even in the absence of topological edge states. In addition, the graphene layer serves as an oxidation barrier, helping protect the intrinsic properties of WTe$_2$ and supporting the potential of this heterostructure for spintronic applications.
\end{abstract}
%\keywords{Graphene, Tungsten ditelluride, Spin Hall Effect, Spin Texture}

\maketitle

Transition metal dichalcogenides (TMDs) exhibit a wide range of electronic properties and topological phases, including topological insulators, semimetals and superconductors.~\cite{Manzeli2017} 
Their structure, typically composed of covalently bonded atomic layers held together by weak van der Waals (vdW) forces, enables the fabrication of heterostructures,\cite{Novoselov2016} which provide an excellent platform for exploring electronic, spin, and topological phenomena in two dimensions (2D), especially via proximity effects.~\cite{Bart2020} For example, interfacing graphene with TMDs can strengthen spin-orbit coupling (SOC) and induce complex spin textures in graphene, enabling spin manipulation and transport despite its otherwise weak spin–orbit interaction.\cite{switch, jose_hugo_garcia} These induced spin textures include not only Rashba and Weyl-like patterns,\cite{Naimer2021, Frank2024, zollner2023} but also Zeeman-type persistent spin textures (PSTs),\cite{yuan2013zeeman} characterized by spin polarization of states that remains unidirectional regardless of momentum over extended regions of the Brillouin zone (BZ).\cite{tao2018PSTenforced, kilic2025universal} The presence of PST can lead to extended spin lifetime and its anisotropy, as confirmed experimentally in graphene/TMD systems.\cite{ghiasi, valenzuela, ingla_aynes}

While persistent spin textures have been predominantly studied in hexagonal TMDs,\cite{Bhowal2020, Jafari2024} materials that typically crystallize in orthorhombic structures, such as WTe$_2$ or MoTe$_2$ have also attracted attention in the context of magnetotransport, efficient charge-to-spin conversion and long-range spin transport.\cite{Tian2023, Anna2024, Vila2021} In particular, monolayer WTe$_2$ manifests a quantum spin Hall effect (QSHE) with spin polarization that is neither orthogonal nor parallel to spin current.\cite{garcia2020canted} This unconventional canted spin configuration reflects a distinct type of PST, different from the purely out-of-plane spin textures typical of hexagonal TMDs. Experimental demonstrations of the QSHE at low temperatures suggest its topological insulator nature,\cite{ wte2edgeconduction6, QSHE100kwte224, Cobden2021, Tan2021} establishing WTe$_2$ as a promising platform for topological spin transport. At the same time, the bulk T$_\mathrm{d}$ phase hosts Weyl points and Fermi arcs, positioning WTe$_2$ as a rare material bridging 2D topological insulators and 3D Weyl semimetal physics.\cite{Mazhar2014} 

Despite its potential, WTe$_2$ is highly sensitive to oxidation under ambient conditions, which complicates device integration.\cite{Ye2016, hou2020oxidation-wte2, Tang2025} However, the variety of spin-orbit-driven phenomena motivates research into strategies for preserving its properties in heterostructures. Graphene, with its chemical stability and atomically thin structure, offers a natural route for protecting other (sensitive) 2D materials without significantly perturbing their electronic structure.\cite{wojciech2025} However, combining hexagonal graphene with rectangular WTe$_2$ is structurally complicated, making such heterostructures less explored and not fully understood. Still, this mismatch creates an interesting configuration to investigate how proximity effects reshape spin textures and topological properties - whether the quantum spin Hall phase can survive symmetry breaking, or whether new spin-related phenomena, such as two-dimensional Weyl semimetal behavior, might emerge.

In this work, we use first-principles density functional theory (DFT) calculations to study the electronic structure and spin texture of graphene/WTe$_2$ heterostructures. By comparing the monolayer WTe$_2$ with its graphene-covered counterpart, we find that the PST - typically enforced by symmetry - survives even in the absence of global crystal symmetries. We attribute this robustness to the presence of local mirror symmetry retained in certain regions of the heterostructure. While the band gap closes upon interfacing with graphene and the system becomes semimetallic, we observe that it hosts a sizable intrinsic spin Hall conductivity, comparable to other 2D materials with strong SOC. These findings show the potential for exploring spin–orbit-driven phenomena in systems with reduced symmetry and the viability of graphene/WTe$_2$ as a platform for spin-based device applications.

\begin{figure*}
  \includegraphics[scale=1]{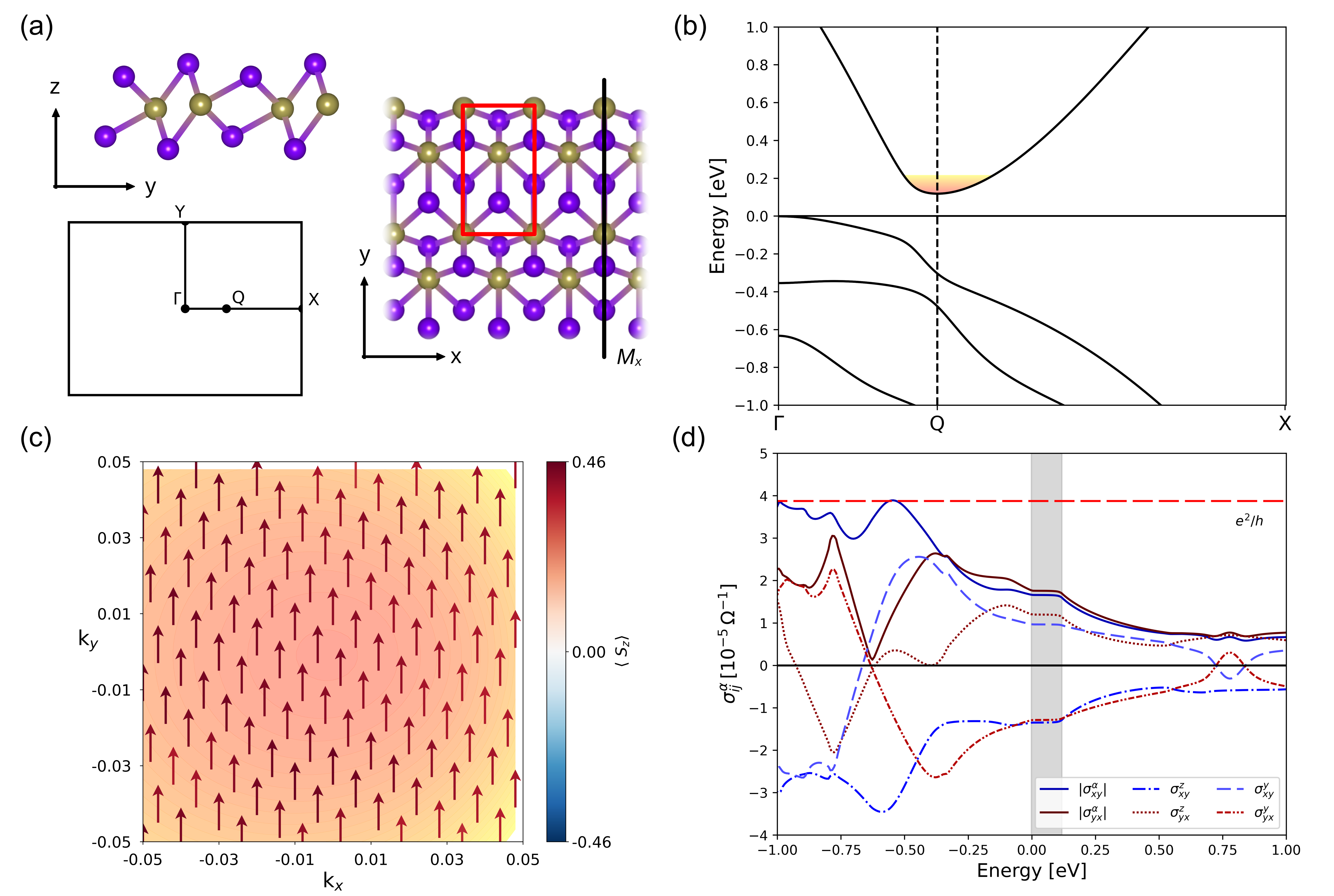}
  \caption{Structure and electronic properties of monolayer WTe$_2$. 
  (a) Side and top views of the primitive cell and the Brillouin zone with high-symmetry points. The mirror plane $M_x$ is indicated by a black line. (b) Band structure along the $\Gamma$–$X$ line. (c) Spin texture of the lowest conduction band around the $Q$ point. The region shown corresponds to the bottom of the electron pocket, highlighted by an orange gradient in (b) that spans an energy range of 100 meV. (d) Calculated spin Hall conductivity. Only the most relevant components exceeding $10^{-6} \ \Omega^{-1}$ are shown. The gray area indicates the band gap.
  }
\end{figure*}

\textit{Monolayer WTe$_2$.} We begin by analyzing the structure and electronic properties of monolayer WTe$_2$. Its crystal structure, shown in Fig.~1a, consists of a rectangular unit cell in which a layer of W atoms is sandwiched between two Te layers. The symmetry of monolayer WTe$_2$ has been reported either as space group (SG) 11, which includes inversion and a mirror plane, or as SG 6, preserving only mirror symmetry. Our structural relaxation shows a slight breaking of inversion symmetry, consistent with SG 6. Both of them contrast with bulk WTe$_2$ in the T\textsubscript{d} phase (SG 31), which, apart from the mirror plane, exhibits glide and screw symmetries related to the translations in the out-of-plane direction. These structural differences lead to markedly distinct electronic properties; while the bulk is a Weyl semimetal with pairs of large hole and electron pockets symmetrically positioned along the $\Gamma$–X direction,\cite{Hyunsoo2019, Jafari2022} the monolayer is a topological insulator with a SOC-induced band gap.\cite{quantumSHETMDS43} Experimental studies report a gap of a few tens of meV, depending on conditions such as temperature, strain, and substrate.\cite{quantumSHEWTe21Tp8, experimentwte2tdmonolayer, Zhao2020, Maximenko2022} 
%This variation arises from the temperature dependence of the band gap, which follows trends observed in other narrow-gap semiconductors, where the gap increases at lower temperatures due to lattice contraction and electron-lattice interactions.\cite{bite2tempbandgap46} 
%Other proposed contributions include incommensurate charge density waves and excitonic interactions.\cite{wte2edgeconduction6,tiWTe2CWD47}

From a computational perspective, DFT calculations using the Perdew–Burke–Ernzerhof (PBE) functional \cite{PBE1996generalized} fail to reproduce the band gap, instead predicting a semimetallic state. To address this, we employed the hybrid Heyd–Scuseria–Ernzerhof (HSE06) functional,\cite{HSE06} which yields a band gap of approximately 118 meV, falling within the range of previously reported values.\cite{zheng2016quantum, garcia2020canted, Maximenko2022} The resulting band structure along the $\Gamma$–X direction is shown in Fig.~1b, featuring the conduction band minimum near the $Q$ point and the valence band maximum at $\Gamma$, again consistent with earlier DFT studies. Figure~1c presents a closer examination of the pocket around $Q$, showing energy eigenvalues within the window indicated by the orange shading in Fig.~1b. Superimposed on the eigenvalues is the spin texture: the in-plane spin components, $S_x$ and $S_y$ are indicated by arrows, while the out-of-plane component ($S_z$) is represented by the color. We observe that the spin polarization has only $S_y$ and $S_z$ components, with $S_x$ vanishing, showcasing a so-called canted persistent spin texture around the $Q$ point, in agreement with previous theoretical predictions.\cite{garcia2020canted} The spin orientation forms an angle of around 62$^\circ$ with respect to the $y$-axis, consistent with earlier calculations and close to the experimental measurements of 50$^\circ$ and 59$^\circ$.\cite{Cobden2021, Tan2021}

The presence of canted PST in the monolayer can be understood in terms of the symmetries. Bulk WTe$_2$ crystallizes in SG 31 (Pmn2$_1$) which is generated by the two mirror symmetries: (i) mirror $M_x=\{m_{100}|0\}$ and (ii) glide mirror $\Tilde{M_y}=\{m_{010}|\frac{1}{2}, 0, \frac{1}{2}\}$. Thus, the $\Gamma-X$ line, which contains $Q$, is left invariant by $\Tilde{M_y}$ and $\mathcal{T}M_x$:

\hspace*{0em}
\begin{minipage}{\dimexpr\linewidth-5em}
\begin{align*}
    \Tilde{M_y}: \Vec{k}\otimes\Vec{\sigma} \rightarrow (k_x, -k_y, k_z)\otimes(-\sigma_x,\sigma_y,-\sigma_z) \\
    \mathcal{T}M_x: \Vec{k}\otimes\Vec{\sigma} \rightarrow (k_x, -k_y, -k_z)\otimes(-\sigma_x,\sigma_y,\sigma_z)
\end{align*}
\end{minipage}
\vspace{1em}

\noindent where $\mathcal{T}$ is the time-reversal operator. $\mathcal{T}M_x$ enforces that the spins of nondegenerate bands to lie on the $yz$-plane, whereas $\Tilde{M_y}$ locks them along the $x$-axis. \cite{kilic2025universal} In contrast to bulk WTe$_2$, which exhibits the true glide mirror symmetry $\Tilde{M_y}$, in monolayer WTe$_2$ this symmetry is broken due to the loss of periodicity along the $z$-axis. With $\Tilde{M_y}$ broken and $M_x$ preserved, the spins are confined to the $yz$-plane, tilting from $y$-axis toward the $z$ direction.

Based on the accurate band structure, we proceed to compute the spin Hall conductivity (SHC) of monolayer WTe$_2$. Because it is known to be a quantum spin Hall insulator, obtaining a finite band gap was essential to capture its topological transport properties. Our results, shown in Fig.~1d, reveal plateaus in the SHC within the gap region, consistent with the quantum spin Hall effect. However, unlike previous theoretical work,\cite{garcia2020canted} which reported a quantized SHC of $2e^2/h$ within the gap, our calculations do not show such quantization. In that study, the quantized value was associated with a spin current polarized along the canted spin direction $\alpha$, with the corresponding SHC defined as $|\sigma^\alpha_{ij}| \equiv \sqrt{(\sigma^y_{ij})^2 + (\sigma^z_{ij})^2}$, where $\alpha$ represents the canted spin polarization direction, and $i,j$ denote the directions of spin and charge currents, respectively. In contrast, our results yield non-quantized values for both $|\sigma^\alpha_{xy}|$ and $|\sigma^\alpha_{yx}|$. Although we observe clear plateaus within the gap, the magnitudes are considerably below the reciprocal von Klitzing constant $e^2/h$.
%In our DFT-based calculations, however, the corresponding SHC within the gap reaches approximately $0.4e^2/h$.

Importantly, a similar lack of quantization in the spin Hall conductivity has been observed in DFT studies of other 2D topological insulators with rectangular lattices, such as WS$_2$ and MoS$_2$, in contrast to systems with hexagonal or square symmetry.\cite{quantizationSHEDFT} The deviation from quantization has been linked to the non-conservation of the spin component corresponding to the spin current’s polarization direction, which can occur, for instance, due to Rashba-like spin-orbit interaction.\cite{quantizationSHEDFT} In our DFT calculations (Fig.~2b), the valence band maximum is located at $\Gamma$, where no PST is present. This contrasts with the previous calculation based on the model,\cite{garcia2020canted} where the valence band maximum was located at the $Q$ point hosting the symmetry-enforced PST. Since the spin Hall conductivity involves integration over the entire Brillouin zone, the absence of PST around $\Gamma$ likely contributes to the reduced, non-quantized SHC observed in our calculations.

\textit{Graphene/WTe$_2$ heterostructure.} We constructed an untwisted graphene/WTe$_2$ heterostructure by matching a $(2\times5)$ supercell of WTe$_2$ with a $(5\times4)$ reconstructed rectangular cell of graphene (see Fig. 2a). The lattice parameters of graphene were set to {2.46} \AA \space and {4.27} \AA, and those of WTe$_2$ to {6.16 \AA\space and 3.42 \AA,} resulting in lattice strains below 2.5\%. The resulting supercell contains 140 atoms, and its space group was identified as $P1$ (SG~1) using the \textsc{Spglib} package.\cite{togo2024spglib} Although the system has no exact symmetries beyond the identity operation, we observe that in certain regions, the mirror symmetry $M_x$ of WTe$_2$ is locally preserved, despite being broken globally. This local symmetry can act as a constraint in momentum space, %potentially enforcing additional degeneracies on the $M_y$-invariant plane. 
enforcing a canted persistent spin texture in a manner similar to that observed in monolayer WTe$_2$.

\begin{figure*}
  \includegraphics[scale=1]{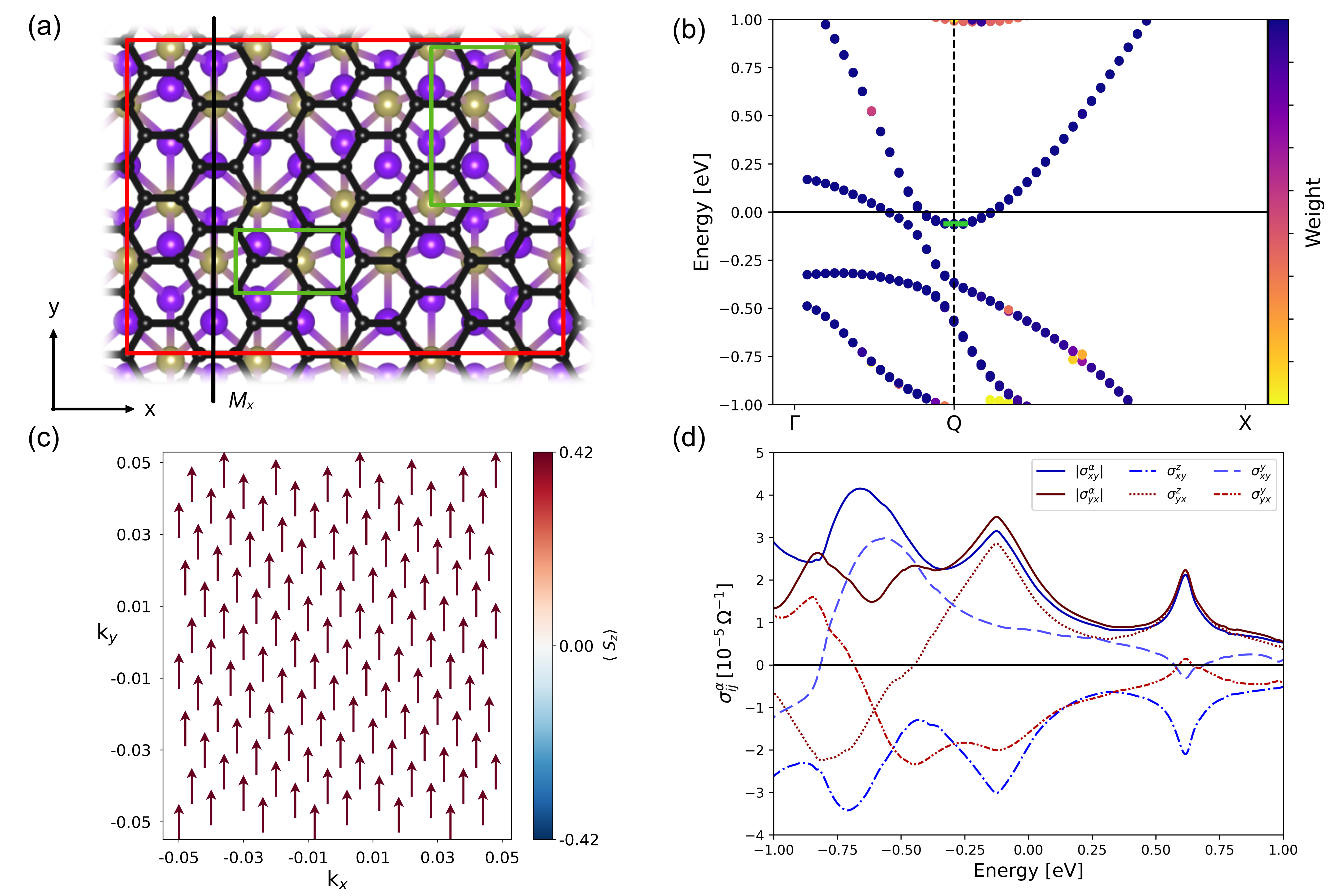}
    \caption{Properties of the graphene/WTe$_2$ heterostructure. (a) Top view; the supercell cell is marked as a red rectangle, and the unit cells of monolayer WTe$_2$ and the rectangular unit cell of graphene are shown as green rectangles. The local mirror symmetry $M_x$ is indicated as a black line.  (b) Band unfolding along the $\Gamma$–$X$ path of WTe$_2$. (c) Spin texture around the $Q$ point, evaluated at an isoenergy of –61 meV (marked by a green line in panel b). (d) Calculated spin Hall conductivity.
}
\end{figure*}

The electronic properties of WTe$_2$ interfaced with graphene are analyzed using band unfolding, a technique that maps the band structure of a supercell onto the BZ of a simpler primitive cell. This allows a direct comparison between the electronic states of the heterostructure and those of the pristine material, providing insight into how interlayer coupling modifies the properties of WTe$_2$. The band structure unfolded onto the WTe$_2$ monolayer is presented in Fig.~2b. Comparing the bands of pristine WTe$_2$ (Fig.~1b) with those of the graphene/WTe$_2$ heterostructure (Fig.~2b), we observe that the conduction band minimum at $Q$ shifts downward by approximately 0.2 eV, and the valence band maximum at $\Gamma$ moves upward by about 0.2 eV, resulting in a semimetallic system. While these band modifications may suggest the conversion into a 2D Weyl semimetal, no band crossings are found in the considered energy window. Instead, we observe an avoided crossing near the $Q$ point at the Fermi level, indicating that graphene does not induce new topological features but only drives the system into a semimetallic state.

Figure 2c shows the spin texture of the conduction band around the $Q$ point at the energy level marked in green in Fig.~2b. It closely resembles that of the pristine monolayer, being primarily composed of $S_y$ and $S_z$ components. The canting angle in the heterostructure is about 62° with respect to the $y$-axis, which is identical to the canting angle in the monolayer WTe$_2$. This robustness of the canted spin orientation, despite the loss of global symmetry, emphasizes the role of weak interaction with graphene and locally preserved mirror symmetry in maintaining the PST. Thus, the local symmetry still plays an important role in locking the spins within the $k_y$–$k_z$ plane, potentially contributing to unconventional (canted) SHE.

%Figure 2c shows the spin texture of the conduction band around the $Q$ point at the energy level marked in green in Fig.~2b. It closely resembles that of the pristine monolayer, being primarily composed of $S_x$ and $S_z$ components, with only a negligible $S_y$ contribution. A noticeable difference from the pristine WTe$_2$ monolayer is that the $S_z$ component, and consequently the canting angle of the spin polarization relative to $k_x$, varies in the immediate neighborhood of the $Q$ point, ranging from zero to \textcolor{red}{64°}, comparable to the canting angle of 62° in monolayer WTe$_2$. \textcolor{red}{The origin of this variation can be explained as ... This demonstrates that PST is largely preserved in the presence of graphene, indicating that the local symmetry still plays an important role, locking the spins within the $k_y$–$k_z$ plane, potentially contributing to canted SHE.}

In the absence of any nontrivial crystal symmetries, all 27 components of the SHC tensor are, in principle, allowed in the graphene/WTe$_2$ heterostructure.\cite{Roy2022} However, the response remains dominated by just four components contributing to $\sigma^\alpha_{xy}$ and $\sigma^\alpha_{yx}$, where $\alpha$ denotes a canted spin direction, similar to the monolayer case. This selectivity shows that the essential charge-to-spin conversion configurations of WTe$_2$ are preserved even after interfacing with graphene.
Although the transition to a semimetal eliminates the possibility of a quantum spin Hall effect, which is reflected by the lack of plateaus in the SHC plots in Fig.~2d, the values of SHC remain substantial, reaching approximately \( 0.5\, e^2/h \) at the Fermi level, very close to the value found in the monolayer within the band gap. Compared to other 2D materials reported in the recent high-throughput study,\cite{Charlier2025} this value surpasses that of many 2D monolayer metals, although higher SHC can still be found in a few systems. Thus, graphene capping preserves not only the spin texture but also the efficiency of spin Hall effect, making the heterostructure competitive within the broader family of 2D materials.

From the perspective of spin-based technology, the efficient charge-to-spin conversion via canted spin Hall effect, especially when reinforced by persistent spin texture, can be harnessed to generate spin–orbit torques and switch magnets with perpendicular magnetic anisotropy.\cite{Vila2021} Even though the topological states are absent in the heterostructure, PST has been shown to support long spin lifetimes in various material families, including monolayer WTe$_2$ where it leads to anisotropic spin relaxation, with significantly extended spin lifetimes along the direction of canted spin polarization.\cite{Ping2024} Beyond improving charge-to-spin conversion characteristics, such anisotropy provides a foundation for exploring new spintronic device architectures based on 2D materials.\cite{slawinska2020ultrathin} While spin lifetime and spin relaxation mechanisms in the graphene/WTe$_2$ heterostructure have yet to be fully characterized, the presence of PST points to promising directions for future theoretical and experimental exploration.

\textit{Conclusion.} We studied the electronic and spin properties of monolayer WTe$_2$ covered by graphene using first-principles calculations. Despite the emergence of semimetallic behavior, the characteristic canted spin texture near the $Q$ point remains largely preserved due to locally retained mirror symmetry and weak interlayer interaction. The heterostructure exhibits canted spin Hall conductivity with magnitudes comparable to other 2D materials with strong SOC, supporting efficient charge-to-spin conversion and potentially enhanced spin lifetimes enabled by the persistent spin texture. Although pristine monolayer WTe$_2$ can host a quantum spin Hall effect at low temperatures, its narrow band gap is extremely fragile, and the material is highly sensitive to oxidation. In contrast, the graphene-covered structure offers environmental stability and maintains spin-related properties under ambient conditions. These advantages highlight the strong potential of graphene/WTe$_2$ heterostructure for efficient generation and transport of spin signals, making it highly relevant for future spin-based technologies.

\textit{Methods.} 
%\textcolor{red}{For the DFT calculations of pristine monolayer WTe$_2$, we used the \textsc{VASP} package. The plane-wave energy cutoff was set to 400~eV, and a $24 \times 24 \times 1$ Monkhorst-Pack $k$-point mesh was employed. The cell dimensions were defined as $[[0, -6.28~\text{\AA}, 0], [3.5~\text{\AA}, 0, 0], [0, 0, 30~\text{\AA}]]$ and subsequently relaxed. The internal atomic coordinates were relaxed until the forces acting on atoms were below $2\ \times  10^{-3}$~eV/\AA. The band gap was obtained using HSE.}  
Our first-principles DFT calculations for the WTe$_2$ monolayer and the graphene/WTe$_2$ heterostructure were performed with the \textsc{Quantum Espresso} package.~\cite{giannozzi2009quantumespresso,giannozzi2017advancedquantumespresso}Electron–ion interactions were treated using the projector augmented-wave (PAW) method, and wavefunctions were expanded in a plane-wave basis with kinetic energy cutoffs of 80 Ry for the monolayer and 60~Ry for the heterostructure. Exchange–correlation functionals were defined within the generalized gradient approximation (GGA),\cite{perdew1996generalizedGGA} using the Perdew–Burke–Ernzerhof (PBE) parametrization for the monolayer and PBEsol for the heterostructure. The band gap of the WTe$_2$ monolayer was obtained using the HSE06 functional and renormalized with a scissors operator at the postprocessing stage. 

For the monolayers, we used the lattice parameters set to {2.46} \AA \space and {4.26} \AA \space for graphene and {6.28} \AA\space and {3.5} \AA \space for WTe$_2$. The graphene/WTe$_2$ heterostructure supercell, constructed using the \textsc{MedeA} software, contained 140 atoms, with cell dimensions given by  
[[12.32~\text{\AA}, 0, 0], [0, 17.08~\text{\AA}, 0], [0, 0, 40~\text{\AA}]]. The corresponding compressed primitive cell of WTe$_2$, extracted from the supercell, had lattice parameters 6.16~\text{\AA} and 3.42~\text{\AA}, and for graphene 2.46~\text{\AA} and 4.27~\text{\AA}. To minimize spurious interactions between periodic images, vacuum spacings of 26~\AA\ (monolayer) and 32~\AA\ (heterostructure) were used. Atomic positions were fully relaxed until residual forces were below $10^{-4}$Ry/Bohr, including van der Waals interactions treated with the DFT-D3 approach.\cite{dft-d3} The equilibrium interlayer separation in the relaxed heterostructure, defined as the average distance between tungsten and carbon atoms, was $d_0 \approx 5.47$~\AA.

The Brillouin zone was sampled using a $24 \times 24 \times 1$ $k$-point mesh for the monolayer and a $4 \times 4 \times 1$ mesh for the heterostructure.  
Spin–orbit coupling (SOC) was included self-consistently in all calculations, excluding during structural relaxations. Band unfolding was carried out using the \textsc{BandUP(py)} code~\cite{medeiros2014effectsbandup,medeiros2015unfoldingbandup,iraola2022irrepbandup}. Spin textures and spin Hall conductivity were obtained using the \textsc{paoflow} package~\cite{PAOFLOW1,PAOFLOW}.  
For the SHC calculations, the PAO Hamiltonians were interpolated to a $96 \times 96 \times 1$ mesh for the monolayer and $24 \times 24 \times 1$ mesh for the heterostructure.

%As an additional check, Chern numbers were independently verified using the \textsc{Z2PACK} code \cite{gresch2017z2pack,soluyanov2011computingz2pack}.

\textit{Acknowledgments.}

This work was financially supported by the National Science Centre, Poland under projects \verb|2018/30/E/ST5/00667| (P.P. and P.D.), \verb|2018/31/B/ST3/02450| (P.J.K.) and implemented as part of the "Smarter, Faster, Better! – Internationalisation of Doctoral Schools at the University of Łódź" project funded by STER NAWA - Internationalisation of Doctoral Schools Programme. J.S. acknowledges the Rosalind Franklin Fellowship from the University of Groningen. J.S. and B.K. acknowledge the Dutch Research Council (NWO) - grant OCENW.M.22.063. The calculations were carried out on the Dutch national e-infrastructure with the support of SURF Cooperative (EINF-8924) and on the Hábrók high-performance computing cluster of the University of Groningen.  

\textit{Data availability.}
All data associated with the results reported in this paper are available from DataverseNL.

%\bibliographystyle{apsrev4-1} 
%\bibliographystyle{abbrvnat}
%\bibliography{lib}

\end{document}